\documentclass[letterpaper,twocolumn,10pt]{article}
\usepackage{usenix-2020-09}

\usepackage{tikz}
\usepackage{amsmath}
\usepackage{comment}
\usepackage{pifont}
\usepackage{algorithm}
\usepackage[noend]{algpseudocode}
\usepackage{xspace}
\usepackage{enumitem}
\usepackage{makecell}
\usepackage{booktabs}
\usepackage{subcaption}
\usepackage{listings}

\usepackage{ulem}
\usepackage{multirow} 

\usepackage{stfloats}

\microtypecontext{spacing=nonfrench}

\newcommand*\circled[1]{\tikz[baseline=(char.base)]{
            \node[shape=circle,fill,inner sep=1pt] (char) {\footnotesize \textcolor{white}{#1}};}}

\newcommand{\name}{KVDirect}

\begin{document}

\date{}

\title{{\name}: Distributed Disaggregated LLM Inference}

\newif\ifshowauthor

\showauthortrue

\ifshowauthor

\author{
{\rm Shiyang Chen}\thanks{This work began during his internship at ByteDance.}\\
Rutgers University
\and
{\rm Rain Jiang}\\
ByteDance
\and
{\rm Dezhi Yu}\\
ByteDance
\and
{\rm Jinlai Xu}\\
ByteDance
\and
{\rm Mengyuan Chao}\\
ByteDance
\and
{\rm Fanlong Meng}\\
ByteDance
\and
{\rm Chenyu Jiang}\\
ByteDance
\and
{\rm Wei Xu}\\
ByteDance
\and
{\rm Hang Liu}\thanks{Corresponding author.}\\
Rutgers University
}

\newcommand{\repo}{\url{https://github.com/TensorDirect/KVDirect}}

\else
\newcommand{\repo}{}
\fi

\maketitle

\begin{abstract}
Large Language Models (LLMs) have become the new foundation for many applications, reshaping human society like a storm. Disaggregated inference, which separates prefill and decode stages, is a promising approach to improving hardware utilization and service quality. However, due to inefficient inter-node communication, existing systems restrict disaggregated inference to a single node, limiting resource allocation flexibility and reducing service capacity. This paper introduces {\name}, which optimizes KV cache transfer to enable a distributed disaggregated LLM inference. {\name} achieves this through the following contributions. First, we propose a novel tensor-centric communication mechanism that reduces the synchronization overhead in traditional distributed GPU systems. Second, we design a custom communication library to support dynamic GPU resource scheduling and efficient KV cache transfer. Third, we introduce a pull-based KV cache transfer strategy that reduces GPU resource idling and improves latency. Finally, we implement {\name} as an open-source LLM inference framework. Our evaluation demonstrates that {\name} reduces per-request latency by 55\% compared to the baseline across diverse workloads under the same resource constraints.

\end{abstract}

\section{Introduction}

Large Language Models (LLMs) are revolutionizing a broad spectrum of applications and pioneering new frontiers in research and industry. Beyond chatbots, LLMs have become integral in diverse fields, fueling innovations from genomic research~\cite{zvyagin2023genslms, zhang2023dnagpt} to molecular design~\cite{mazuz2023molecule,cho2023iupacgpt, bagal2021molgpt}. These models are also increasingly embedded in everyday activities such as news reading~\cite{chern2023factool}, search engines~\cite{xiong2024search, ma2024crafting}, and personalized recommendation systems~\cite{petrov2023generative}, reshaping how we interact with information and making profound impacts across all sectors of our daily life.

LLM serves the end users through LLM inference, which contains two steps: prefill and decode: (i) The \textit{prefill} stage processes the prompt and generates the first token in the response. Following it, (ii) the \textit{decode} stage continues generating the response. Specifically, the decode stage consists of multiple iterations, each generating one token based on the previously generated text. To speed up the inference, the intermediate embeddings of all prompt tokens are stored in the KV cache during the prefill stage. During the decode stage, the model computes the embedding of the newest token and reuses the existing embeddings from the KV cache. The LLM service quality is measured
by two key metrics: {the Time To First Token} (TTFT), which is the duration of the (i) prefill stage, and {the Time Between Tokens} (TBT), which represents the average time taken to
generate a response token (except the first token) for (ii) decode step. TTFT and TBT jointly determine the per-request latency.

Noticing the difference between prefill and decode stages~\cite{strati2024d, zhong2024distserve, patel2024splitwise}, disaggregated LLM inference is introduced to serve these two stages separately for better service quality and resource utilization. This approach separates the inference workload between two independent workers. The prefill worker computes the KV cache for all tokens in the prompt, and the computed KV cache is then transferred to the decode worker for subsequent generation. 
As a result, when receiving a new request with a long prompt, the slow prefill computation of the new request will not interrupt the generation iterations of the decode phase from the existing ongoing requests, reducing the TBT. Moreover, one can allocate computation resources to prefill and decode workers differently according to the length of the prompts and responses, which will meet the TTFT and TBT requirements with more efficient resource utilization.

We observe a significant limitation faced by the existing disaggregated LLM inference efforts~\cite{patel2024splitwise,zhong2024distserve}, that is, they only work with a single machine, in part because they would like to resort to fast intra-node NV-Link to transfer the KV cache between prefill and decode workers. However, limiting disaggregated inference on a single machine brings in the following two serious problems:

Firstly, sharing the same node for prefill and decode workers limits the overall service capacity. For example, the prefill and decode worker sharing an 8-GPU node leads to 160 GB of memory for each worker. The system hence can accommodate 83 8K-token prompts under 7B model. When the model size increases to 70B, the prompt length reduces to 700 to maintain the same concurrency level. 
DistServe~\cite{zhong2024distserve} discussed the possibility of splitting the model layers across multiple nodes to reduce the per-node memory requirements of a worker. However, this approach introduces significant communication overhead for intermediate results. Furthermore, this design can only achieve limited scalability because all workers must use at least one GPU in any node. Therefore, DistServe failed to distribute the disaggregated LLM inference.

Secondly, colocating prefill and decode workers restricts resource allocation flexibility. For instance, when scaling the system to support longer contexts, both prefill and decode workers are added with the new node, even if additional decode capacity is unnecessary. Adapting to varying workloads further requires adjusting the ratio of prefill to decode workers within a node. As a result, existing systems rely on performance models~\cite{zhong2024distserve, strati2024d} or simulators~\cite{patel2024splitwise, agrawal2024vidur} to estimate performance under different configurations. However, workload fluctuations over time lead to clusters with inconsistent prefill-to-decode worker ratios, which complicates the scheduling and resource management.

This paper advocates distributing the prefill and decode workers to enable disaggregated LLM inference. However, we notice that adopting the mainstream message-passing-based paradigm to enable distributed disaggregated LLM inference would face severe performance degradation. Below, we summarize the three research challenges faced by the naive message-passing-based paradigms: 
(i) Message-passing-based design incurs many rounds of waiting, synchronization, and data movements, resulting in only 13.6\% of the communication being effective.  
(ii) Message-passing-based libraries are suitable for transferring a large chunk of contiguous memory space, which is not the case for KV cache in LLM. 
(iii) Existing KV cache transfer design holds the precious GPU memory for an exceedingly long period without using it, which restricts the total number of active requests that the decode worker can handle, incurring queuing. (Please refer to Section~\ref{sec:motivation} for detailed analysis).

To address these challenges, we develop {\name}, which optimizes the KV cache transfer to enable distributed disaggregated LLM inference. The contributions of this paper are summarized as follows:

\begin{itemize}
    \item We propose a novel tensor-centric communication mechanism that is tailored for KV cache transfer in disaggregated LLM inference, which avoids multiple round-trip communication and synchronization overheads.

    \item We implement the communication mechanism based on GPU RDMA with the inter-node KV cache transfer. It supports the establishment of dynamic connections with efficient data transfer designs.

    \item We propose the KV cache transfer using pull-mode to let the decode worker read data from the prefill worker. This improves the GPU resource utilization and performance under high QPS.
    
    \item We implement and evaluate our distributed disaggregated LLM inference framework {\name}. This framework is open-sourced and will be available upon publication.
\end{itemize}

This paper is structured as follows.  Section~\ref{sec:background} provides background about disaggregated LLM inference and GPU RDMA.  Section~\ref{sec:motivation} identifies and discusses the key roadblocks hindering distributed disaggregated LLM inference. Section~\ref{sec:system} presents the design and implementation of {\name}. Section~\ref{sec:evaluation} evaluates its performance. Section~\ref{sec:related} reviews related work in the field. We discuss the future work in Section~\ref{sec:discussion}, and conclude in Section~\ref{sec:conclusion}.

\section{Background}
\label{sec:background}

\begin{figure*}[ht]
    \centering
    \includegraphics[width=\textwidth,page=1]{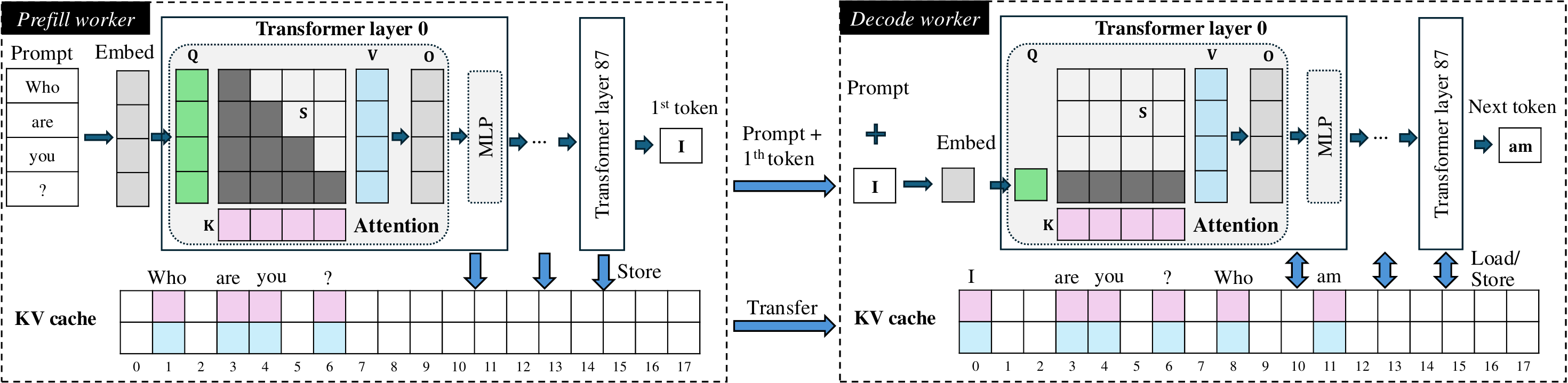}
    \caption{The workflow of disaggregated LLM inference with an emphasis on KV cache.
    \vspace{-.2in}
    }
    \label{fig:infer}
\end{figure*}

\subsection{LLM Inference}
\autoref{fig:infer} illustrates disaggregated LLM inference with a toy example.  
Particularly, each LLM layer (i.e., transformer layer) comprises an attention module and an MLP module. In the example, a prompt containing four tokens is converted into four token embeddings. During attention computation, these embeddings are used to derive the \textbf{Q}, \textbf{K}, and \textbf{V} tensors, and the \textbf{O} tensor is computed as follows:
\begin{align}    
    \textbf{S}_{ij} &= \frac{\exp(\mathbf{Q}_i\mathbf{K}_j^T)}{\sum^{k\leq i}_{k=0} \exp(\mathbf{Q}_i\mathbf{K}^T_k)}, \\
    \textbf{O}_i &= \sum^{k\leq i}_{k=0} \textbf{S}_{ik} \mathbf{V}_k,
\end{align}
where the i$^{th}$ token is attended with all previous tokens. The resultant \textbf{O} tensor is passed to the MLP module to finish this layer computation. After processing through all model layers, four token embeddings are computed and translated to output tokens. According to the definition of the auto-aggressive model, each output token is the next possible token after its previous tokens in prompts. In the generative tasks, only the final preceding token is needed as it represents the predicted next token of the whole prompt. The computation is performed iteratively until all tokens are generated. 

The \textbf{K} and \textbf{V} tensors from previously computed tokens can be cached to optimize computation. As shown in the example, the output token depends on the last row in the attention matrix \textbf{S}, allowing the reuse of the \textbf{K} and \textbf{V} tensors from earlier tokens. 

LLM inference contains two stages: \textit{prefill} and \textit{decode}. The prefill stage computes the attention for every pair of tokens in the prompt, and the \textbf{K} and \textbf{V} tensors are stored in the KV cache. Note that the KV cache is organized into discrete blocks to accommodate variable prompt and response lengths~\cite{kwon2023efficient}. In the example, the four tokens in the prompt are stored in blocks 1, 3, 4, and 6 in the prefill worker.
During the decode stage: the \textbf{K} and \textbf{V} tensors of all previous tokens are used to generate the newest token. Finally, the KV cache of the newly generated token is stored in the KV cache of the decode worker. With the help of the KV cache, the computation of token generation in LLM is reduced from $O(L^2)$ to $O(L)$ at the cost of $O(L)$ memory usage.

\subsection{Why Disaggregated LLM Inference?}

\begin{figure}[h]
    \centering
    \includegraphics[width=0.8\columnwidth]{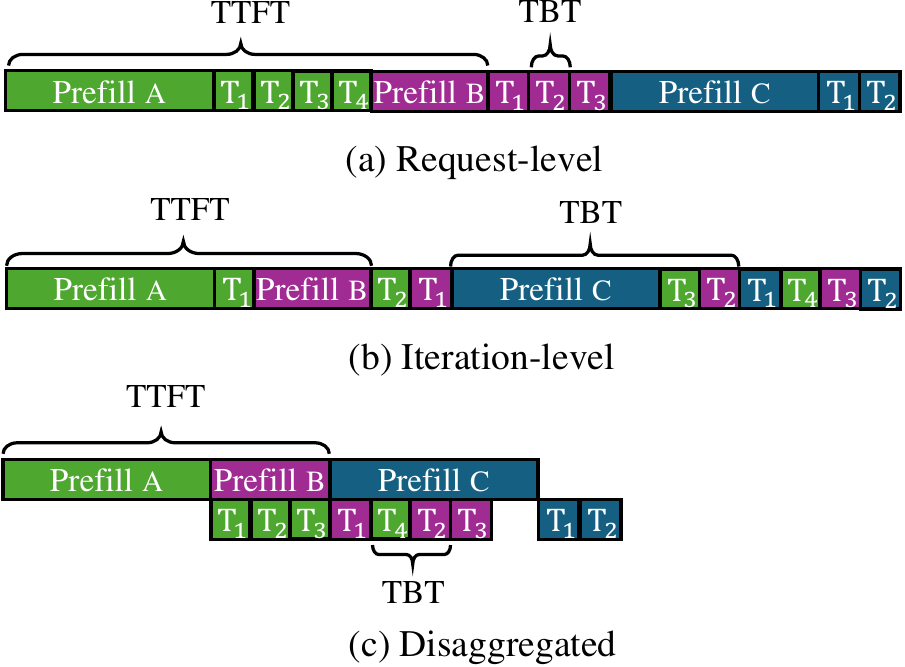}
    \caption{An example with three LLM inference scheduling approaches according to Splitwise~\cite{patel2024splitwise}. Here, we assume prompts A, B, and C arrive in a serial order.}
    \label{fig:schedule}
\end{figure}

\noindent
\autoref{fig:schedule} illustrates three popular scheduling strategies in LLM inference, each aiming to improve service quality. \autoref{fig:schedule}(a) depicts the naive request-level scheduling, which requires finishing one request before processing the next. In this approach, request B has to wait for the complete service of request A. This leads to a long TTFT for request B.

Iteration-level scheduling, as shown in \autoref{fig:schedule}(b), interleaves the computation iterations of different requests to enhance efficiency~\cite{yu2022orca}. For example, once the first decoding iteration (T$_1$) of request A is completed, the prefill for request B can start, effectively reducing the TTFT for request B. However, this strategy may increase the TBT for request B because the prefill stage of request C could be served before T$_2$. This generally leads to unstable service quality.

Disaggregated LLM inference addresses both aforementioned concerns. As shown in \autoref{fig:schedule}(c), one worker is dedicated to the prefill tasks while the other is for decode tasks. This design enjoys a short TTFT and stable TBT. 
Of note, since disaggregated LLM inference uses extra GPUs, so we should compare the latency using scaled per-node QPS.

\subsection{GPU RDMA}
GPU RDMA (Remote Direct Memory Access) is a pivotal technology that enables direct data transfers between GPUs across different nodes, bypassing the CPU. Traditional data transfers move data from the GPU to the CPU and then pass it to the NIC for transmission. Then, the procedure is reversed on the remote node to reach the destination GPU. This data path introduces additional latency and PCIe bottlenecks, especially in multi-GPU systems. GPU RDMA, in contrast, allows the data to be transferred from GPU to NIC with Direct Memory Access (DMA), minimizing CPU and PCIe involvement and significantly improving bandwidth efficiency.

In RDMA-based communication, data transfers occur by posting RDMA verbs. The process starts with establishing a queue pair on the NIC, consisting of a send queue and a receive queue. Before any transfer, a contiguous memory block is registered as a Memory Region (MR), enabling direct access by the NICs. The application then transfers data by posting RDMA verbs to the NIC and monitoring completion queues for status updates. For example, when sending a message, the initiator posts a \textit{read} verb, specifying the size of the message and its offset in the MR. Following a one-sided manner, the responder is not needed to process this communication, reducing the overhead. Additionally, the \textit{send} verbs can also be used for message-passing communication, but it requires the responder to post a \textit{receive} verb to specify the memory to be written.

\section{Motivations}
\label{sec:motivation}

\begin{figure}[ht]
    \centering
    \includegraphics[width=\columnwidth,page=1]{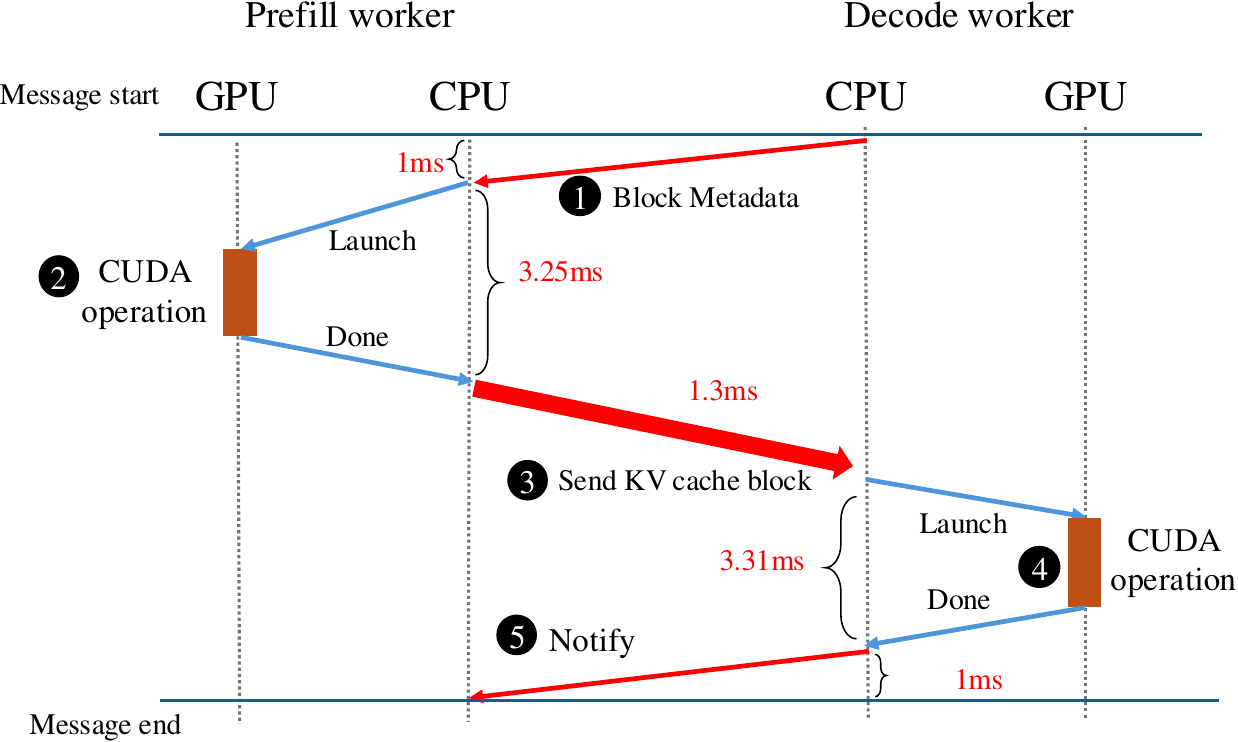}
    \caption{The message-based KV cache transfer with 4KB block size, where the blue and red arrows represent the communication over PCIe and network, respectively.
    \vspace{-.2in}
    }
    \label{fig:message}
\end{figure}

\subsection*{Motivation \#1: Traditional distributed GPU communication design is inefficient.}

\autoref{fig:message} illustrates a typical workflow of transferring \textit{a single KV cache block} if the state-of-the-art (i.e., DistServe~\cite{zhong2024distserve} and Splitwise~\cite{patel2024splitwise}) would expand their disaggregated LLM inference into distributed settings with any existing communication strategy: Step \circled{1} (1ms): The decode worker initiates the communication and sends block metadata via RPC, such as the block ID and the request ID, to the prefill worker. 
Step \circled{2} (3.25ms): When the prefill worker receives the metadata, it launches GPU operations, including the communication kernel, and copies the data to the buffer.
Step \circled{3} (1.3ms): 
After finishing the GPU operations, the CPU synchronizes with the GPU and operates the NIC to send the KV cache block.  
Step \circled{4} (3.31ms): After receiving the data, the decode worker launches corresponding GPU operations to place the data in the KV cache of the decode worker. 
Step \circled{5} (1ms): Once the block is processed, the decode worker notifies the prefill worker of completion and starts the next iteration.

Traditional distributed GPU communication design is time-consuming and will take most of the disaggregated LLM inference. First, the inefficiency of KV cache transfer arises primarily from waiting and synchronization overheads. In the example, the actual KV cache transfer time (step \circled{3}) accounts for only 13.2\% of the total transfer duration. Further, transferring the KV cache from prefill workers to decode workers is, in fact, a major bottleneck in disaggregated LLM inference: This inefficiency stems from two factors: {the massive size of the KV cache and its fragmented storage. For instance, in 70B models, a single request with a 16K-token prompt generates 640 MB of KV cache, split into 2048 disjoint 4KB blocks for a single GPU out of eight. In this case, the prefill computation of this request would only take 0.9 seconds, while transferring it costs 2.7 seconds.}

\subsection*{Motivation \#2: The state-of-the-art high-performance communication libraries are unsuitable for KV cache transfer.}

High-performance communication libraries, such as NCCL~\cite{NCCL}, UCX~\cite{shamis2015ucx}, and MSCCL++~\cite{mscclpp}, all follow a message-passing-based mechanism, which is unsuitable for KV cache transfers from the following two aspects:

Firstly, these libraries cannot support the changes of participating GPU devices during execution. Optimized for bulky collective communication, where large amounts of data are exchanged among all ranks in the system, message-passing-based frameworks require a communication graph to be built at the initialization stage, i.e., MPI\_Init. This permits the library to derive an optimal transfer scheme along different paths simultaneously to maximize bandwidth. However, this static setup rules out the possibility of dynamically adding or removing GPUs to the service. In the event of that, the disaggregated inference service must be restarted to rebuild the communication graph~\cite{torch_elastic}. 

Second, existing libraries only support the transfer of contiguous memory. To send noncontiguous KV cache blocks for a prompt, users must either launch separate transfers for each block or gather all blocks into a single user buffer before sending. Both approaches introduce additional overhead due to repeated kernel launches and CPU-GPU synchronization. Libraries like NVSHMEM~\cite{NVSHMEM} may resolve this issue by providing synchronous shared memory across GPUs. However, the decode worker dynamically allocates new blocks, which should not be synchronized with the prefill worker, during response generation, but NVSHMEM will force this synchronization. That said, NVSHMEM adds unnecessary network traffic and increases memory usage on the prefill worker.

\begin{figure}
    \centering
    \includegraphics[width=0.9\columnwidth]{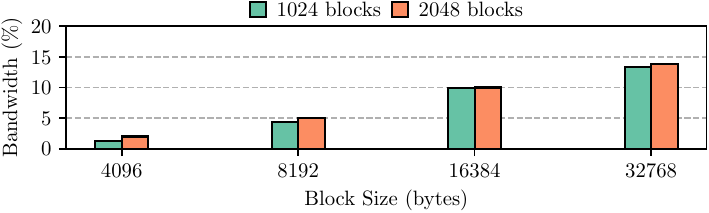}
    \caption{The achieved bandwidth of UCX message-sending.
    \vspace{-.2in}
    }
    \label{fig:bandwidth}
\end{figure}

\autoref{fig:bandwidth} illustrates the achieved bandwidth when transferring 1024 and 2048 blocks using UCX~\cite{shamis2015ucx}. The results show that transferring 4KB blocks utilizes only 1.8\% of the available bandwidth, while larger block sizes up to 32KB achieve a maximum utilization of just 13.6\%. This inefficiency is compounded by the synchronization overhead and the inability of small block transfers to saturate the NIC, further reducing effective bandwidth. Moreover, transferring fewer blocks exacerbates bandwidth underutilization. For instance, sending 1024 4KB blocks (corresponding to 8K tokens) achieves 40\% less bandwidth than transferring 2048 blocks of the same size.

We also noticed that several RDMA-based efforts aim to address the challenges inherent in message-passing, but these solutions fail to resolve the issues specific to KV cache transfer. On the one hand, many approaches focus on CPU-only communication scenarios~\cite{wang2023replicating, li2023flor,wei2023no,wei2018deconstructing}. However, in KV cache transfer, GPU-CPU synchronization is a critical bottleneck, as the GPU must wait for kernel completion and data copying to the CPU. On the other hand, techniques such as zero-side RDMA~\cite{jasny2024zero} advocate fully bypassing the CPU by allowing the GPU to manage the NIC directly. While this eliminates CPU involvement in data transfer, it is not practical for LLM inference, where tasks like KV cache allocation and token decoding are required to run on the CPU.

\begin{figure*}[ht]
    \centering
    \includegraphics[width=\textwidth]{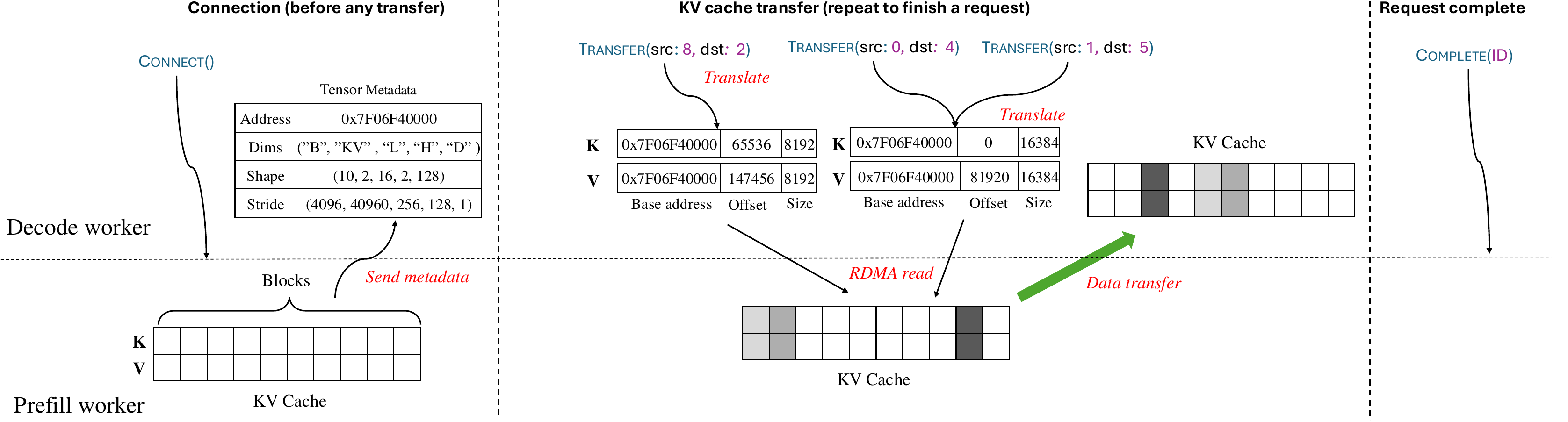}
    \vspace{-.3in}
    \caption{
    An example of {\name}'s tensor communication.
    \vspace{-.05in}
    }
    \label{fig:tensor}
\end{figure*}

\subsection*{Motivation \#3: GPU resources are held but idling for too long in the existing designs. }

Recent endeavors of disaggregated LLM inference favor the ``push-mode'' in KV cache transfer to mitigate the communication bottleneck. Specifically, once the prefill worker completes one layer, it can push the associated KV cache to the decode worker, hiding the transfer overhead with the subsequent computations. However, to enable this approach, all KV cache blocks for a request must be pre-allocated on both the prefill and decode workers, even though only one layer is transferred at a time. Incremental, on-demand allocation of KV cache blocks is impractical because it risks deadlock when GPU memory is fully utilized. The cause is as follows: if multiple requests deplete the GPU memory on the decode worker, but none have sufficient memory to proceed, they end up waiting indefinitely for other requests to release memory.

Allocating all the memory resources for the decode worker to wait for the completion of the prefill worker leads to holding a large volume of precious GPU memory resources without using them for too long. First, the KV cache memory consumption is substantial. For example, on a node with 8 GPUs (40 GB each) serving a 70B model, only 34 requests with 16K-token prompts can be accommodated concurrently. Therefore, during the prefill computation, when a large chunk of the memory is held for transfer but not used, this significantly reduces the GPU resource utilization.

\begin{figure}[ht]
    \centering
    \includegraphics[width=0.85\columnwidth]{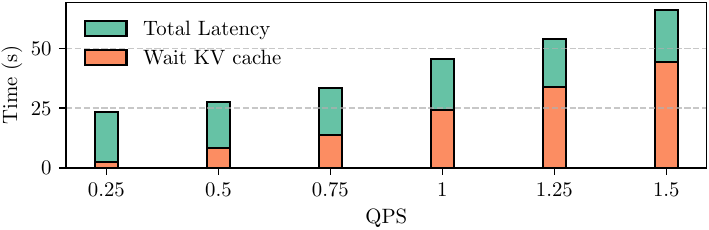}
    \caption{The latency of 16K-token requests with 70B model.}
    \vspace{-0.2in}
    \label{fig:queue_kv}
\end{figure}

Second, when the memory of any worker is fully utilized, the system becomes unavailable to process new requests, leading to significant latency increases. \autoref{fig:queue_kv} shows the per-request latency for 16K-token prompts with a 70B model as the query per second (QPS) ranging from 0.25 to 1.5. The latency increases sharply from 23 seconds to 68 seconds when QPS reaches 2. The primary bottleneck is the waiting for KV cache allocation. At high QPS, when the decode time exceeds the prefill time, the running requests fully occupy the decode worker’s resources, leaving no available memory for new requests. As a result, new requests cannot reserve memory on the decode worker, preventing their prefill computation from starting, even if the prefill worker is idle. Additionally, as QPS increases, the queuing time grows, further exacerbating the waiting time for new requests. At a QPS of 1.5, this bottleneck contributes to 65\% of the total latency.

\section{{\name} Design and Implementation}
\label{sec:system}

\begin{figure*}[ht]
    \centering
    \includegraphics[width=\textwidth]{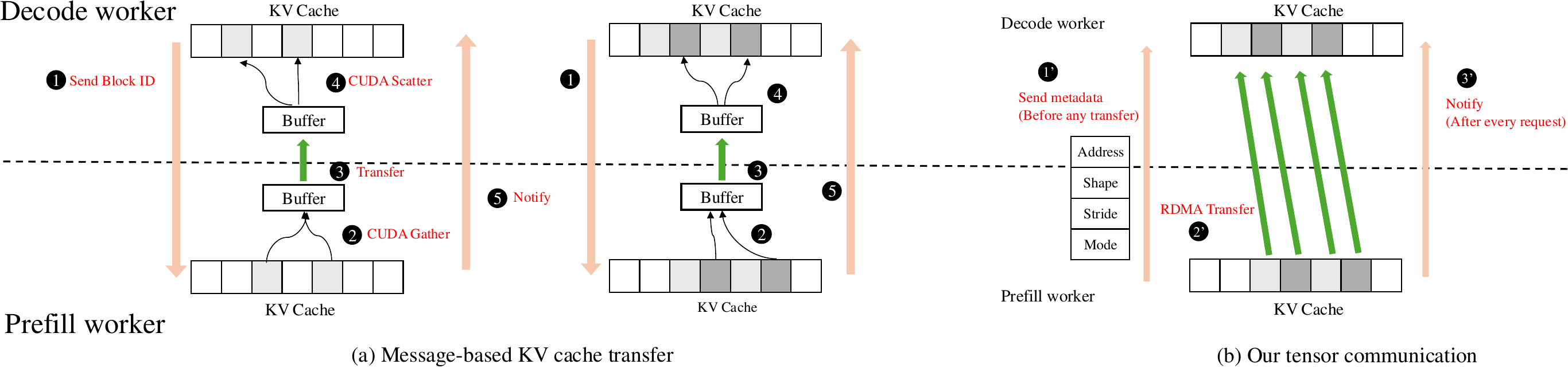}
    \caption{An example of KV cache transfer, where a request with 4 blocks is sent. (a) The message-based KV cache transfer. The four blocks are sent in two iterations. (b) {\name} KV cache transfer.
    \vspace{-.1in}   
    }
    \label{fig:diff}
\end{figure*}

\subsection{{\name} Communication Design}

\autoref{fig:tensor} depicts our {\name} communication mechanism, which largely eliminates the multiple round-trip communication and synchronization overheads (Motivation \# 1). 

The mechanism consists of three key operations: \textsc{Connect()}, \textsc{Transfer()}, and \textsc{Complete()}. First, the decode worker connects to the prefill worker with the metadata using \textsc{Connect()} to initialize the connection. Subsequently, for each cache block, the decode worker fetches the associated KV cache blocks using \textsc{Transfer()}. Typically, each request transfers thousands of blocks. After transferring all blocks belonging to one request, one \textsc{Complete()} is invoked to notify the prefill worker. In short, \textit{compared to traditional design Motivation \# 1, which requires per-block metadata exchange and notification, {\name} reduces that to a negligible frequency. }

In \textsc{Connect()}, the decode worker connects with the prefill worker through a handshake process. Once the connection is established, the prefill worker sends the metadata of every tensor to the decode worker.
This metadata includes the tensor’s \textit{Address}, \textit{Dims}, \textit{Shape}, and \textit{Stride}.

\autoref{fig:tensor} exemplifies these metadata: 
(i) \textit{Address} is the base address of the tensor used in RDMA communication, i.e., 0x7F06F40000. 
(ii) \textit{Dims} describes the intention of each dimension in this 5-dimensional KV cache tensor. These five dimensions specify the memory layout by the state-of-the-art attention kernels~\cite{dao2023flashattention2,xFormers}. In this case, the five dimensions are ordered as ``B'', ``KV'', ``L'', ``H'' and ``D'', a.k.a., cache[B][KV][L][H][D]. They are the \# blocks per KV cache, the \textbf{K} and \textbf{V} tensor for this block, \# tokens per block, \# heads per token, and the dimension of a head. Of note, one can also define a different order of these five dimensions corresponding to the usage situation. 
(iii) \textit{Shape} is an instance of Dim. Particularly, B=10, KV=2, L=16, H=2, and D=128.
(iv) The order of \textit{Stride} follows the definition as \textit{Dims}, and the number refers to the distance between two elements of the specific dimension in memory. In the example, the \textbf{K} tensors of every block are placed contiguously in the memory, so the distance between two \textbf{K} tensors is the size of a sub-tensor (16, 2, 128), which is 4096. Besides, the \textbf{K} and \textbf{V} tensors of each block is separated by 10 sub-tensors so the stride is 40960.

The decode worker invokes \textsc{Transfer()} with a remote prefill worker block ID as the source and a local block ID as the destination to receive blocks from the prefill worker during inference. The remote block ID will be translated into RDMA transaction based on the tensor metadata, specifying the remote memory location of the block. In particular, the block 8 consists of the \textbf{K} and \textbf{V}, indexed as cache[8][0][0][0][0] and cache[8][1][0][0][0], respectively. To get the starting offset of each tensor, we can perform a dot-product between the index and the stride as 
\begin{align*}
    (8, 0, 0, 0, 0) \cdot (4096, 40960, 256, 128, 1)^T \times 2B &= 65536B, \\
    (8, 1, 0, 0, 0) \cdot (4096, 40960, 256, 128, 1)^T \times 2B &= 147453 B,
\end{align*}
where we use bfloat16 as the data type.
Then, we compute the size of a continuous memory space to be transferred that can cover the {``L''}, {``H''}, and {``D''} dimensions. We find the dimension with the largest stride, {``L''}, and then multiply its shape with the stride.
\begin{align*}
    16 \times 128 \times 2B = 8192B.
\end{align*}
As a result, the two tensors in block 8 are stored as two disjoint 8192 B memory spaces, so it incurs two RDMA transactions. 
For blocks 0 and 1, the offset of their \textbf{K} tensors are 0 and 8192. Since each block is 8129 B, the two blocks are adjacent in memory. Therefore, the blocks can be coalesced as one 16384 B RDMA transaction to increase the bandwidth.

Once all blocks for a request have been transferred, the decode worker invokes \textsc{Complete()} to send the corresponding request ID to the prefill worker. It then notifies the inference engine to start computing this request. When the prefill worker receives the \textsc{Complete()} message, it notifies the inference engine to release the KV cache block for this request.

\autoref{fig:diff} illustrates the example of communication patterns for traditional message-based and our approach in {\name}. Existing message-based KV cache transfer are NCCL~\cite{NCCL}, MSCCL++~\cite{mscclpp}, and UCX~\cite{shamis2015ucx}. 
In \autoref{fig:diff}(a), four blocks from a request are sent from prefill to decode worker. In the message-based approach, both workers allocate a communication buffer that can hold two blocks. The decode worker first sends the desired block ID to the prefill worker at step \circled{1}. The prefill worker then gathers the blocks to a communication buffer using a CUDA kernel at step \circled{2}, and send the buffer content to the decode worker at step \circled{3}. Once the decode worker receives the data in the buffer, it launches a CUDA kernel to scatter the blocks at step \circled{4}. Afterward, the workers start another communication for the remaining two blocks. With {\name} communication mechanism shown in \autoref{fig:diff}(b), the prefill worker sends its metadata to the decode worker at step \circled{1'} so that the following block transfer does not need the involvement from the prefill worker but the decode worker can compute the memory location to access. As a result, the four blocks are sent as four RDMA transactions without other operations in between.

\begin{figure*}[ht]
    \centering
    \includegraphics[page=1,width=\textwidth]{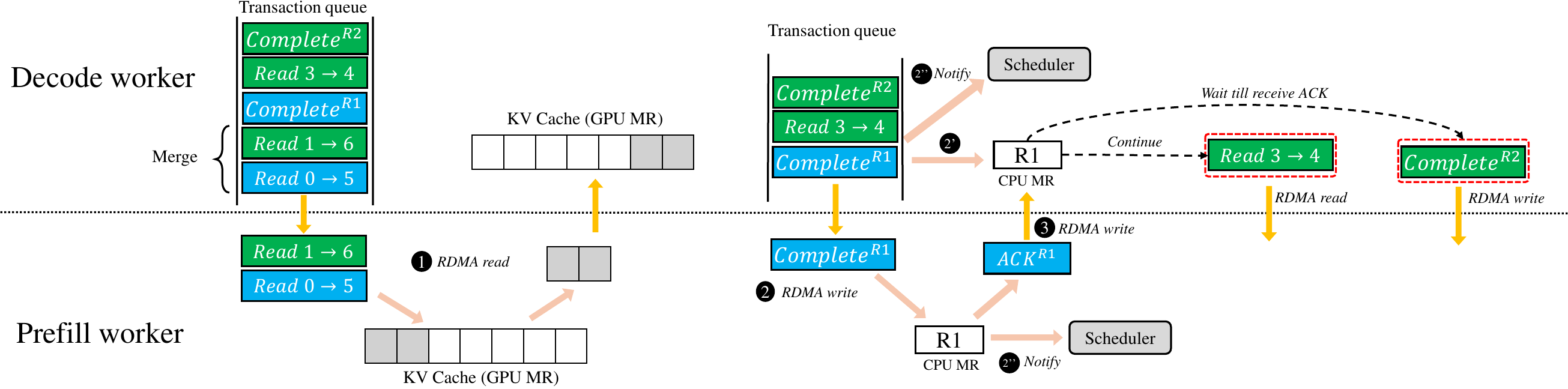}
    \caption{An example of the KV cache communication. There are two requests, R1 (blue) and R2 (green). 
    }
    \label{fig:queue}
\end{figure*}

\subsection{{\name} Communication System}

This section presents the implementation of the aforementioned three key operations: \textsc{Connect()}, \textsc{Transfer()}, and \textsc{Complete()} within the distributed GPU context.

\begin{figure}[h]
    \centering
    \includegraphics[width=\columnwidth]{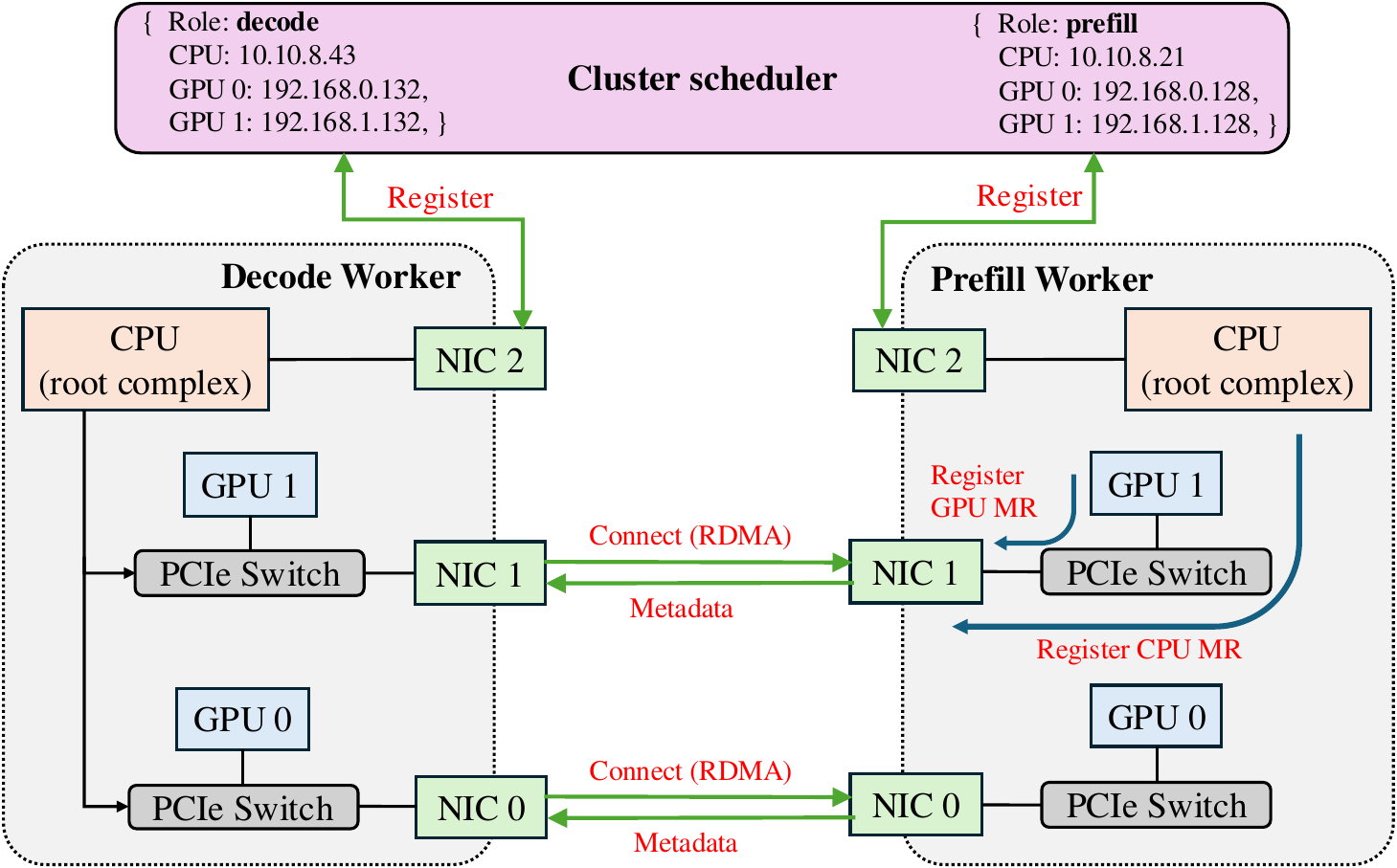}
    \caption{Implementation of \textsc{Connect()} with two machines, each of which features two GPUs and three NICs. 
    }
    \label{fig:connection}
\end{figure}

\textbf{Connection establishment.} 
The connection is established between the NICs of the decode worker and the prefill worker. We pair each GPU with the nearby NIC to implement affinity and bypass communication through the root complex and corresponding latency.
As shown in \autoref{fig:connection}, a worker is equipped with two GPUs and three NICs, following a multi-rail architecture~\cite{coll2003using}. For instance, NIC 0 is paired with GPU 0, which shares the same PCIe switch.

We rely on a cluster scheduler to add and remove workers in an inference cluster (termed cluster hereafter). The cluster scheduler maintains a list of the addresses for active workers. To avoid the single-point failure of the scheduler, the decode worker maintains the connection of all active prefill workers. 

We perform three steps of adding a worker: (i) A user initiates the addition of a worker to this cluster. We connect the worker to the scheduler using the CPU NIC and register its metadata. As shown in the example, the metadata includes the role of the worker, the CPU and two GPUs, and their associated network addresses (e.g., GPU 0: 192.168.0.132 for decode worker). 
(ii) The scheduler broadcasts the new worker's metadata to every other worker in the cluster. If the new worker is a prefill worker, a running decode worker will automatically connect to the new worker based on the NIC address of the newly added worker. In this setup, GPU 0 on both workers is connected via NIC 0, and GPU 1 similarly uses NIC 1. Of note, GPU $i$ of a decode worker can only connect with GPU $i$ of a prefill worker (as shown in \autoref{fig:connection}). 
This is because the typical network architecture of the data center connects the GPUs with different IDs to different spine switches to reduce the network traffic~\cite{wang2024rail}. 
(iii)
Finally, the decode worker connects the prefill worker with the \textsc{Connect()} operations. For every GPU NIC, the worker registers the memory of the KV cache as the GPU MR. Besides, a block of CPU memory is registered to every NIC as the CPU MR to exchange other metadata. 
Once connected, the prefill worker stores the tensor metadata (see \autoref{fig:tensor}) to its CPU MR and posts RDMA \textit{send}. The decoder worker posts RDMA \textit{receive} to receive the tensor metadata via its CPU MR, which completes the connection.

\textbf{Tensor communication.}
The tensor communication of {\name} centers around the transaction queue. Each \textsc{Transfer()} and \textsc{Connect()} operation posts a transaction into the queue, respectively. As the example shown in \autoref{fig:queue}, requests $R1$ (blue) and $R2$ (green) are being queued. $R1$ issues a read transaction reading from the remote block 0 to the local block 5, and then issues a completion transaction marking the end of the requests. Similarly, $R2$ issues two read transactions and one completion transaction. {\name} ensures the \textsc{Complete()} is always invoked after \textsc{Transfer()} operations within one request, while transactions from different requests might be issued out-of-order. 

{\name} processes the transactions from the queue based on the types. Each read transaction posts an RDMA \textit{read} to the NIC, and the prefill worker responds to the blocks to the KV cache on the decode worker (step \circled{1}). The completion transaction sends the request ID to the prefill worker, signaling the completion of the request. The request ID is first copied to the CPU MR (step \circled{2'}) and then sent to the remote CPU MR using RDMA \textit{write} (step \circled{2}). Then, the decode worker notifies its scheduler to start computing, and the prefill worker notifies to release the KV cache blocks (step \circled{2''}).

To mitigate the inefficiency of small blocks and increase the bandwidth, {\name} coalesces read transactions to form larger blocks. Specifically, {\name} pops all the read transactions in order until the first completion transaction for the coalescing opportunity. In the example, the transaction \textit{Read $0 \xrightarrow{} 5$} from $R1$ and \textit{Read $1 \xrightarrow{} 6$} from $R2$ can be merged because both the local and remote blocks for the two transactions are adjacent. Generally, we compute the $\text{Offset} / \text{Size}$ for the local and remote memory location of every transaction. \textit{A group of transactions can be merged only when the results of both remote and local locations are contiguous.}
Then, we post the RDMA \textit{read} for every merged or unmerged transaction without block to maximize bandwidth. We find that the coalescing opportunity is plentiful, especially for long prompts, because of less fragmentation.  

We let the completion messages block each other to prevent write-after-write conflicts in the CPU MR. Since the RDMA \textit{write} is one-sided, if the completion messages of $R1$ and $R2$ are sent without blocking, whichever finishes later might overwrite the preceding \textsc{Complete()} message in the CPU MR, leading to memory leakage concerns. 
Therefore, we use the mechanism of prefill worker sending an ACK message to achieve synchronous \textsc{Complete()} (step \circled{3}).
Of note, the read transactions are still asynchronous and are not blocked by the ACK message; while waiting for the ACK of $R1$, the decode worker can continue transferring blocks for $R2$.

\subsection{Push-mode v.s. Pull-mode}

\begin{figure}[ht]
    \centering
    \includegraphics[width=\columnwidth]{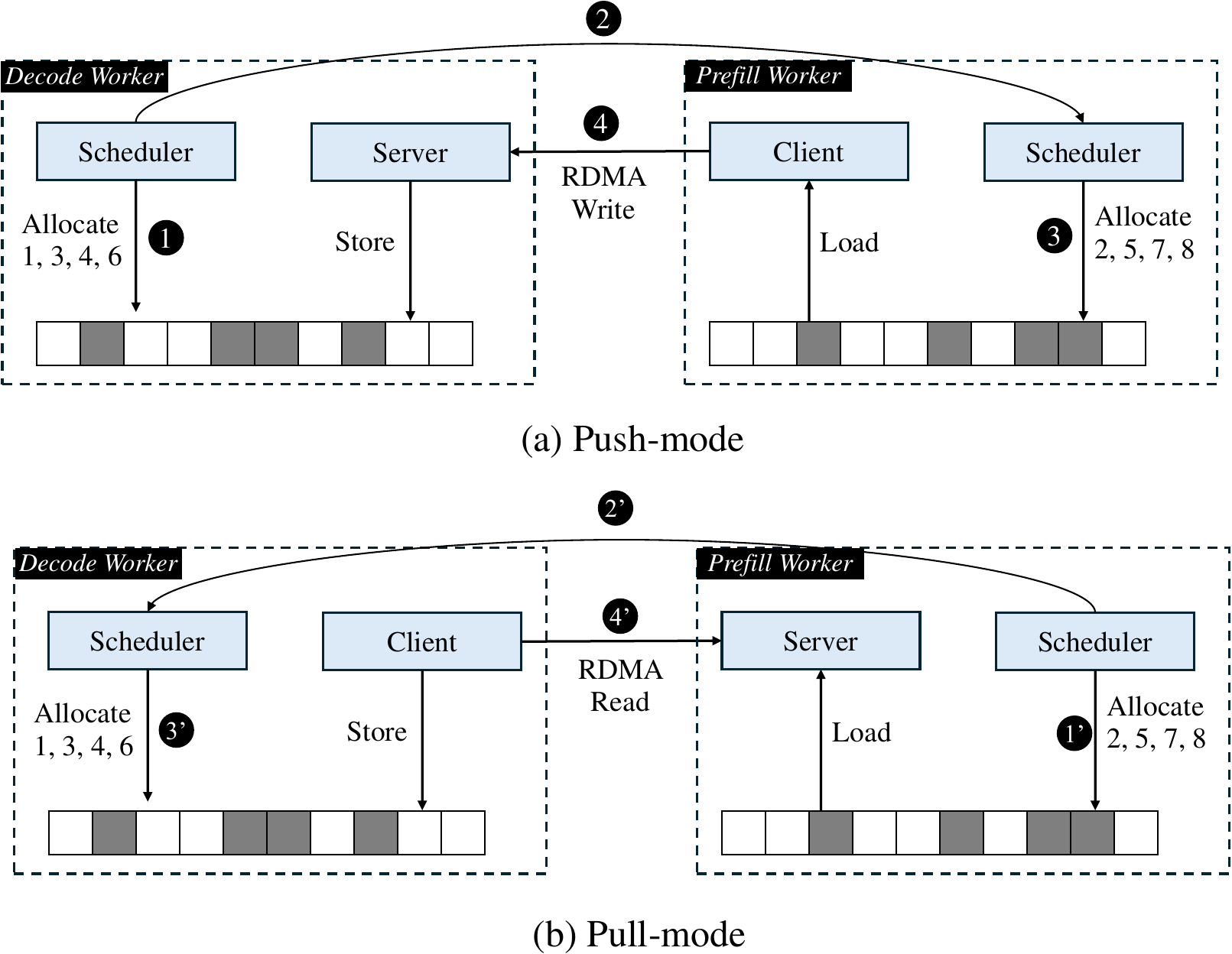}
    \caption{KV cache transfer in push-mode vs. pull-mode.}
    \label{fig:push_pull}
\end{figure}

\noindent
\autoref{fig:push_pull} explains two potential types of transfer patterns in {\name}, i.e., push-mode and pull-mode. First, in push-mode, the prefill worker pushes the KV cache blocks to the decode worker in four steps: At step \circled{1}, the decode worker allocates 4 blocks for the prompt and sends the prompt and the block IDs to the prefill worker as step \circled{2}. The prefill worker allocates the blocks for the prompt at step \circled{3}. Once the prefill worker has finished one layer, it writes the filled blocks to the decode worker based on the received block IDs (\circled{4}) in a pipelined manner.

The pull-mode instead lets the decode worker read from the prefill worker in four steps: At step \circled{1'}, the prefill worker allocates the blocks for the prompt and then performs the prefill computation. After all layers are completed, the block IDs are sent to the decode worker at step \circled{2'}. Correspondingly, at step \circled{3'}, the decode worker allocates the blocks. At step \circled{4'}, the decode worker reads the blocks from the prefill worker via RDMA. Of note, pull-mode performs KV cache reads for all layers in a single shot.

\begin{figure}[ht]
    \centering
    \includegraphics[width=\columnwidth]{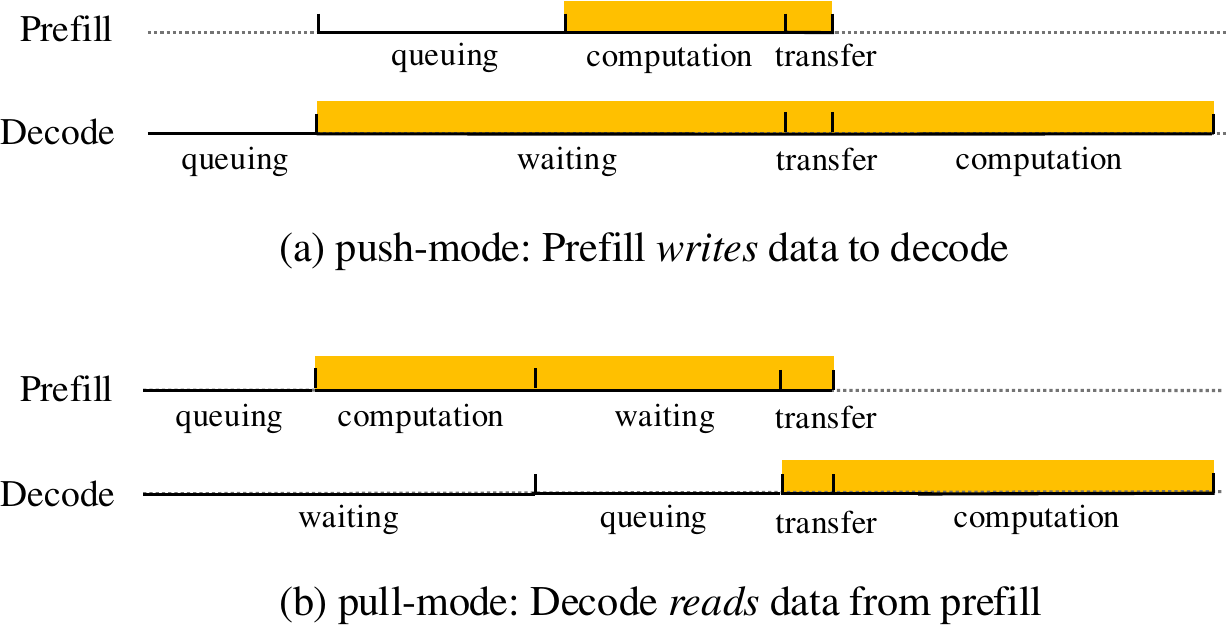}
    \caption{The KV cache lifetime for push-mode and pull-mode, where the shade represents the KV cache in use.
    }
    \label{fig:lifetime}
\end{figure}

\autoref{fig:lifetime} illustrates the KV cache lifetime for push-mode and pull-mode. 
In push-mode, in \autoref{fig:lifetime}(a), the KV cache lifetime on the prefill worker consists of the computation time and transfer time. On the decode worker, this lifetime encompasses all the time spent waiting for the prefill worker to finish, in addition to the decode computation time.
In push-mode, see \autoref{fig:lifetime}(b), the KV cache lifetime on the prefill worker includes computation time, transfer time, and queuing time on the decode worker. The decode worker lifetime in pull-mode only includes transfer and computation times. Note that the queuing time only reflects the wait for KV cache allocation, while the time spent waiting for the scheduler to issue the computation is included in the computation time.

{\name} defaults to pull-mode for its more efficient resource utilization due to shorter KV cache lifetime on the decode worker. In contrast, the push-mode incurs long KV cache lifetime on the decode worker, limiting its number of concurrent requests. Moreover, the prefill worker is idling when requests are queuing for KV cache allocation on the decode worker in push-mode. Conversely, the pull-mode allows the processing on the prefill worker to be in parallel with the KV cache queuing on the decode worker, leading to higher overall system utilization.

{\name} does not opt for pipelined communication for three reasons. 
First, pipelining communications and computations are complex, which can be error-prone and hard to tune for some benefits. 
Second, in pull-mode, once the prefill worker completes a layer, it must notify the decode worker and wait for KV cache allocation on the decode worker for that layer. This waiting time can be substantial, meaning that the actual transfer often occurs after the entire computation. 
Third, with the optimizations in {\name}, the transfer time is negligible compared with the other time-consuming processes, so the benefits of pipelining are small.

\section{Experiments}
\label{sec:evaluation}

\subsection{Experiments Setup.}
\noindent\textbf{Models}
We use the models with randomized weights to demonstrate the performance of {\name}. In particular, we use \textit{Mistral-Large-Instruct-2407}~\cite{mistral123B} for testing, which has 123B parameters and Grouped Query Attention~\cite{ainslie2023gqa} with 8 KV-heads. Therefore, each token incurs 352 KB of memory for the KV cache.

\vspace{.1in}
\noindent\textbf{Datasets.}
We evaluate {\name} using two real-world datasets. The arXiv dataset~\cite{cohan-etal-2018-discourse} contains 216K papers from arXiv. In this evaluation, we use the main content in the prompt and instruct the LLM to summarize the paper. This task has long prompts and relatively short responses, where the average prompt and response lengths are 40,642 and 241.
ShareGPT~\cite{cobbe2021gsm8k} is a collection of user-shared conversations with ChatGPT. The model will generate the last response based on previous chat history. This task has shorter prompts and longer responses, where the average prompt and response lengths are 20,471 and 2,328. We use the Poisson process to simulate the request arrival.

\vspace{.1in}
\noindent\textbf{Metrics.}
We evaluate the inference performance using three metrics: total latency, TTFT, and TBT. Total latency is the end-to-end time from the receipt of a request to the completion of the response. The measured TTFT includes the prefill time and the waiting time for the KV cache. TBT is measured 

\vspace{.1in}
\noindent\textbf{System setting.}
The experiments {\name} are conducted on up to 4 H100 machines, each of which equips 8 H100-80G GPUs, and each GPU is associated with a 400 Gbps RDMA NIC.
We use vLLM v0.5.3 for LLM computation and request scheduling with PyTorch 2.3.0 and CUDA 12.4. 

\begin{figure*}[ht]
\begin{tabular}{c}  

    \subfloat[{Increasing the number of decoder workers and the response length with different prompt lengths.}]{
        \hspace{-.2in}
        \includegraphics[width=\linewidth]{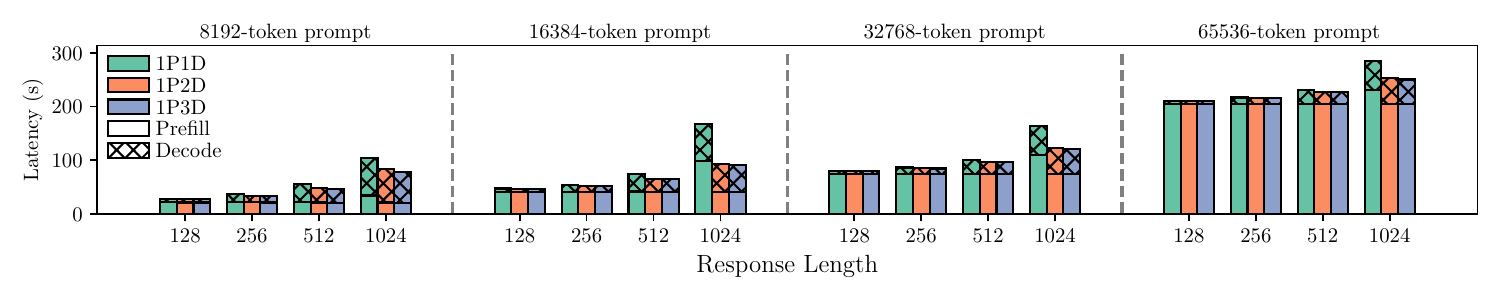}
    }
    \\
    \subfloat[{Increasing the number of prefill workers and the prompt length with different response lengths.}]{
        \hspace{-.2in} 
        \includegraphics[width=\linewidth]{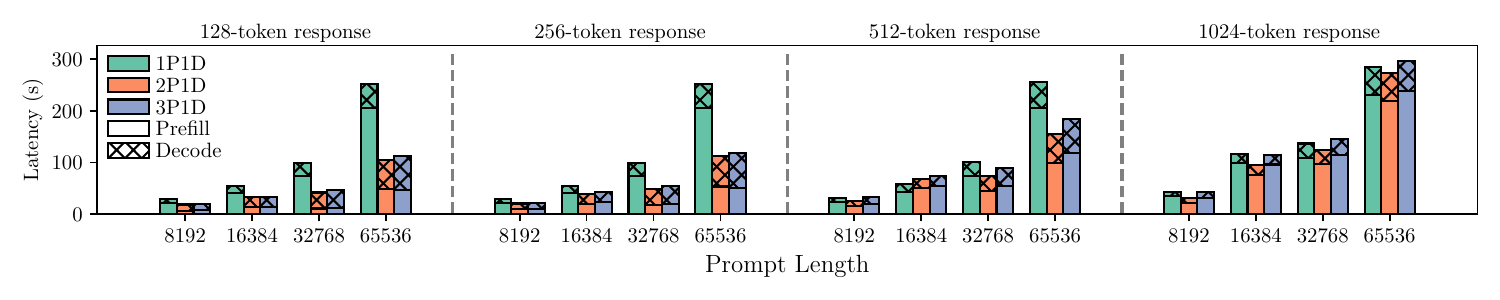}
    } 
\end{tabular}
\caption{{The per-request latency of different cluster configurations of {\name}, where each bar consists of the prefill and decode stage. Note that the prefill stage includes the prefill time and waiting time for the KV cache. }}
\label{fig:eva:scale}
\end{figure*}

\subsection{Inference Performance}
\subsubsection{{\name} Overall Benefits}

\begin{figure}[ht]
    \centering
    \includegraphics[width=\linewidth]{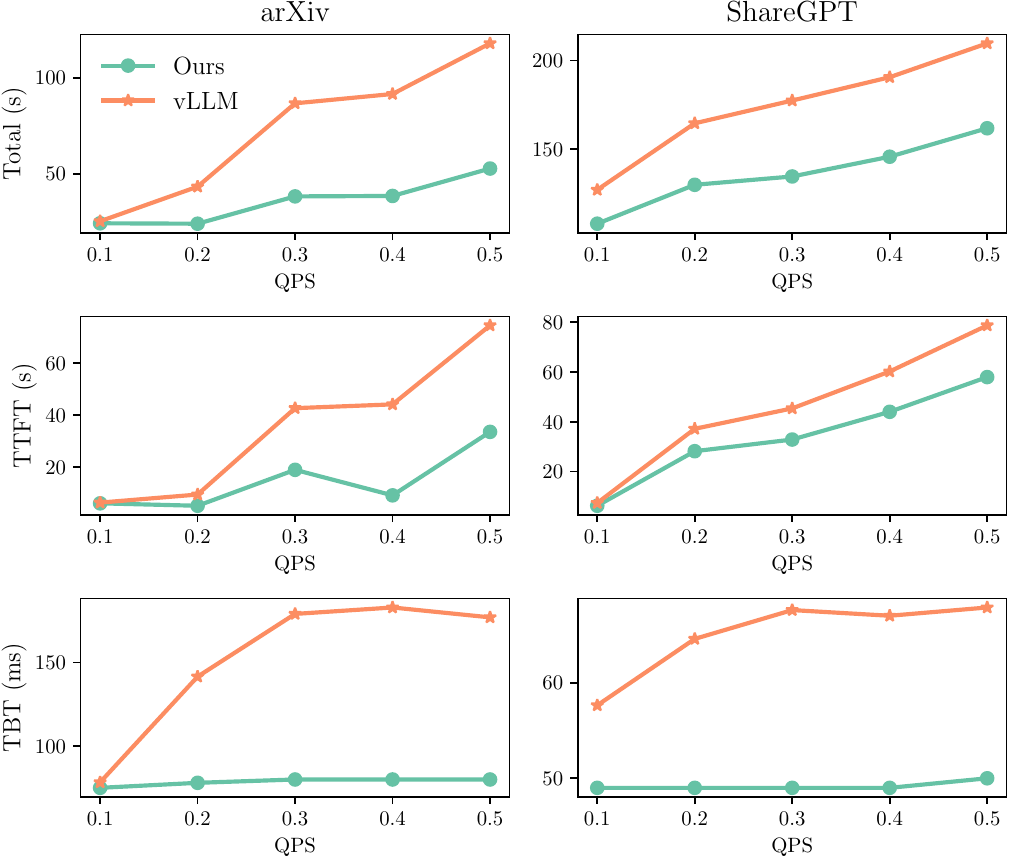}
    \caption{Benchmark results of P90 per-request latency including total latency, TTFT, and TBT (lower is better). 
    \vspace{-.2in}
    }
    \label{fig:eva:benchmark}
\end{figure}

\autoref{fig:eva:benchmark} illustrates the per-request inference latency for {\name} (1 prefill worker, 1 decode worker) and vLLM across different datasets, with each column of graphs representing a dataset and each row representing three metrics. The actual QPS of vLLM is divided by 2 for fair comparison. 

{\name} reduces the per-request latency by 55\% and 24\%, for two datasets respectively. The majority of the speedup is attributed to improvements in TBT, which constitutes over 60\% of the total latency. For vLLM, as the QPS increases, the resource contention of concurrent prefill and decode computations leads to a TBT increase of up to 2.2$\times$ and a TTFT increase of up to 12.3 $\times$. 
In contrast, the TBT of {\name} remains low and the TTFT increases slower than that of vLLM as QPS fluctuations. Of note, the vLLM scheduler prioritizes prefill computation. Consequently, for datasets like arXiv with shorter responses, vLLM can maintain lower TTFT at low QPS levels. However, as QPS increases, the prefill computation interrupts the generation process, rapidly escalating TBT. At high QPS levels, the TBT of vLLM becomes stable because the extended decode time of ongoing requests delays their completion, which in turn depletes the available resources. This delay prevents new requests from allocating the KV cache, thereby rapidly increasing the TTFT.

It is important to note that there is no publicly available \textit{distributed} LLM inference system directly comparable to {\name}. Splitwise~\cite{patel2024splitwise} released a prototype patch on vLLM.~\footnote{\url{https://github.com/vllm-project/vllm/pull/2809}} However, it only supports intra-node communication, and the prefill and decode workers cannot run concurrently. DistServe~\cite{zhong2024distserve} also cannot support inter-node communication.
Mooncake~\cite{Mooncake} releases a prototype.~\footnote{\url{https://github.com/kvcache-ai/Mooncake/}} However, the current implementation does not support tensor-parallelism due to the conflict with NCCL.~\footnote{\url{https://github.com/vllm-project/vllm/pull/10502}}
Other systems like DéjàVu~\cite{strati2024d} are also unavailable for comparison.

\subsubsection{Cluster Configuration Study}

\autoref{fig:eva:scale} evaluates the impact of different cluster configurations on performance. We test {\name} with four prompt lengths, ranging from 8,192 to 65,536 tokens. Each prompt length is paired with four response lengths, from 128 to 1,024 tokens. For instance, {the workload designated as \underline{8192-512} comprises a prompt of 8192 tokens paired with a response of 512 tokens.} 
To ensure all workers are adequately loaded and reveal the effects of queuing and computation, we set the QPS to 8, 4, 1, and 0.6 for the four prompt lengths, respectively.

\autoref{fig:eva:scale}(a) shows the effect of increasing the number of decode workers. Overall, adding more decode workers mitigates the latency increase from longer responses. Increasing the number of decode workers reduces decode stage time by lowering resource contention between requests. For example, the TBT for the \underline{8192-1024} workload decreases from 67 ms to 55 ms, and for the \underline{65536-1024} workload, it decreases from 52 ms to 45 ms when the number of decode workers is increased from 1 to 3. Notably, longer prompts exhibit smaller TBT due to fewer concurrent requests. Furthermore, adding decode workers also reduces the prefill stage time when requests must stay on prefill workers to wait for available KV cache on decode workers. 
For long response lengths such as 1024, the reduction is significant as the decode stage dominates. For the prompt length ranging from 8,192 to 65,536, using 3 decode workers leads to $38\%$, $58\%$, $33\%$, and $11\%$ reduction on prefill time compared to 1 decode workers, respectively.

\autoref{fig:eva:scale}(b) highlights the impact of increasing the number of prefill workers. When the number of prefill workers is increased from 1 to 2, prefill time is reduced by $2.34\times$, $1.74\times$, $3.73\times$, and $4.04\times$ for prompt lengths ranging from 8,192 to 65,536, respectively. Prefill time generally decreases with more prefill workers, especially for long prompts, where the quadratically increasing computation load saturates the GPUs.
Note that in the case of the 8,192 prompt length, the high QPS leads to a long queuing time, resulting in a higher speedup. Interestingly, when increasing the number of prefill workers from 2 to 3, an increase in per-request latency is observed. Specifically, prefill time rises with response length, and both prefill and decode times are longer than those of the 2P1D configuration. This increase is due to higher prefill capacity creating more concurrent requests on the decode worker, which in turn intensifies resource contention, increasing overall per-request decode time. Consequently, prefill time is further prolonged due to queuing for KV cache on the decode worker as response length grows.

\begin{figure}[ht]
    \centering
    \includegraphics[width=\columnwidth]{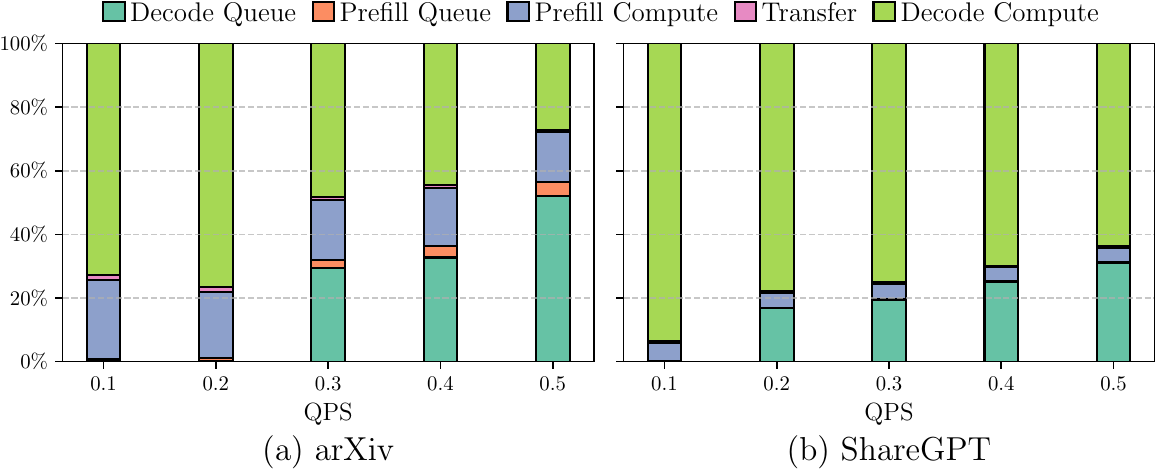}
    \vspace{-.1in}
    \caption{Latency breakdown of arXiv and ShareGPT.
    }
    \label{fig:eva:breakdown}
\end{figure}

\subsubsection{Latency Breakdown}

\autoref{fig:eva:breakdown} provides a breakdown of latency across different stages in the processing lifecycle of a request, segmented into prefill queuing, prefill computation, KV cache transfer, decoding queuing, and decoding computation for the arXiv and ShareGPT datasets. Notably, the transfer time is minimal, accounting for only 1.1\% and 0.5\% of the total latency for the two datasets, respectively. This negligible impact is a testament to the effectiveness of our optimizations in the transfer process.

Moreover, most of the inference latency is dominated by decode-related activities, encompassing both decode computation and decode queuing. Typically, decode computation time surpasses prefill computation time, establishing the decode worker as the bottleneck, especially as the QPS increases. For instance, as QPS rises to 0.5, the proportion of time spent in decode queuing escalates to 52\% and 30\% for each dataset, respectively. Consequently, this increase in decode queuing time significantly contributes to the overall escalation in total latency as QPS increases.

\subsection{Techniques Evaluation}

\begin{figure}[ht]
    \centering
    \includegraphics[width=.9\columnwidth]{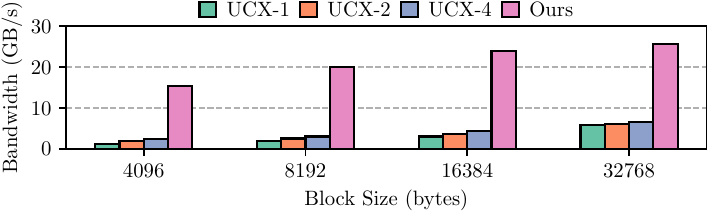}
    \caption{Bandwidth utilization of {\name} and UCX with different number of connections. 
    }
    \label{fig:eva:transport}
\end{figure}
\textbf{{\name} communication.} \autoref{fig:eva:transport} shows the bandwidth utilization of {\name}. In particular, we implement the message-based KV cache transfer using UCX~\cite{ucx-py} as the baseline. We transfer 1024 blocks between 2 H100 GPUs on two nodes via 400 Gbps NIC. Besides, we create multiple connections for UCX to increase its throughput further. The Figure shows that {\name} achieves 22.23 GB/s bandwidth across different block sizes on average, while UCX only uses 4.05 GB/s when using 4 connections. 
As the block size increases, the utilization of both {\name} and UCX increases because of the reduced CPU overhead ratio. Besides, using more connections in UCX improves the bandwidth, especially for small blocks. The reason is that different connections form the pipeline to hide overheads, such as CPU-GPU synchronization and buffer copying. However, when the block size is large, using more connections leads to fewer benefits because the communication kernel becomes the bottleneck. It is noteworthy that further increasing the connection may result in a crash or data loss because of the extensive resource usage required for UCX to maintain a connection.

\begin{figure}
    \centering
    \includegraphics[width=\columnwidth]{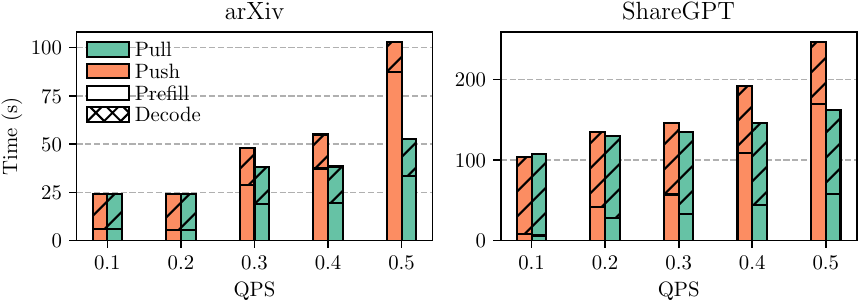}
    \caption{The results of the comparison between pull-mode and push-mode.}
    \label{fig:eva:push-pull}
\end{figure}

\textbf{Pull-mode v.s. Push-mode.} 
\autoref{fig:eva:push-pull} illustrates the comparison between pull-mode and push-mode, demonstrating that pull-mode leads to an average speedup of 25.5\% over push-mode. This improvement primarily stems from reduced waiting times at the decode worker, which are recorded as the prefill time. As the QPS increases, resource idling results in an additional $1.6\times$ increase in queuing time for the KV cache on the decode worker. Additionally, reserving KV cache on the decode worker in push-mode limits the number of concurrent ongoing requests. This constraint results in fewer interruptions between the generations of two consecutive tokens of the same request, thereby reducing the TBT by 5.6\% and 14.4\% across the datasets, respectively. However, despite these improvements in TBT, the total latency still increases in push-mode, primarily because waiting for KV cache becomes a significant bottleneck when the QPS is high.

\begin{figure}[ht]
    \centering
    \includegraphics[width=\columnwidth]{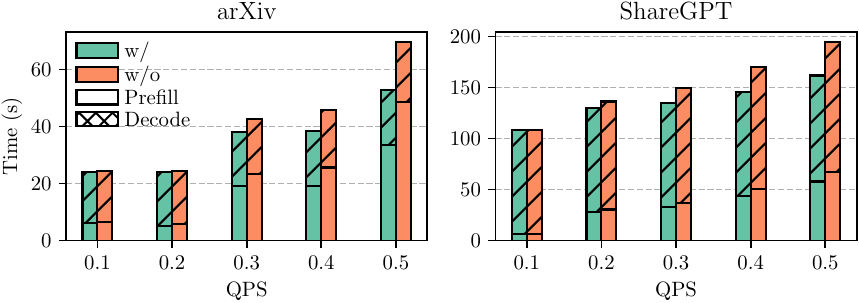}
    \caption{The results of the comparison between with and without block coalescing.}
    \label{fig:eva:block}
\end{figure}

\textbf{Block coalescing.} 
\autoref{fig:eva:block} presents the results of our evaluation on the effectiveness of block coalescing, which demonstrates average speedups of $1.13\times$ and $1.03\times$ for two datasets, respectively. When QPS reaches 0.5, the potential for coalescing increases as more requests are batched together during one prefill computation, resulting in further speedups of $1.32\times$ and $1.07\times$ for the two datasets. Notably, the arXiv dataset achieves a larger speedup due to two key factors: longer prompts, which decrease fragmentation by enabling more contiguous block coalescing, and shorter responses, which require fewer new blocks on the decode worker, further reducing fragmentation and easing memory allocation.

\section{Related Work}
\label{sec:related}

Recent years have witnessed a surge of optimization efforts in LLM inference~\cite{holmes2024deepspeed,agrawal2023sarathi,qiu2024efficient,yu2022orca}. These optimizations aim to improve the efficiency of the model computation and the hardware utilization. This section focuses on the work about disaggregated inference and optimizations of KV cache.

\textbf{Pioneers of disaggregated LLM inference} explore the separation of prefill and decode stages to optimize resource usage. For example, Splitwise~\cite{patel2024splitwise} independently schedules resources for the prefill and decode stages and mixes these stages on the same node to handle burst requests, offering flexibility in balancing resource allocation. However, the KV cache transfer emerges as a bottleneck. To address this, DéjàVu~\cite{strati2024d} introduces a pipeline strategy that transfers the KV cache layer-by-layer and prompt-by-prompt. DistServe~\cite{zhong2024distserve} colocates prefill and decode workers on the same node to utilize high-speed interconnects. Compared with our {\name}, these systems use the message-based KV cache communication, such as MPI in DéjàVu and MSCCL++ in Splitwise, which suffer from low bandwidth utilization. Although DistServe avoids message-sending overhead by implementing KV cache transfers as memory copies, it restricts all workers to a single node, so the number of GPUs per node limits the number of total workers.

\textbf{KV cache reusing} focuses on optimizing shared prefixes across prompts to reduce redundant computations by reusing the KV cache. Hydragen~\cite{juravsky2024hydragen} proposes an implementation of attention to leverage the shared prefix within a batch of prompts. Similarly, SGLang~\cite{zheng2023efficiently} leverages a Radix tree structure to store the KV cache from different prompts. Memserve~\cite{hu2024memserve} and Mooncake~\cite{Mooncake} further extend this to preserve the KV cache in dedicated storage in disaggregated inference, enabling KV cache offloading and buffering at scale. Memserve, for instance, designs an API to index and access KV cache in a distributed memory pool, and Mooncake introduces a locality-aware scheduling strategy to balance workloads and maximize cache hits. In comparison, our {\name} can be used to improve the KV cache movement in the prefix cache. Compared with {\name}, these systems employ CPU-centric communication where the KV cache is transferred between CPUs and then moved to GPUs. As a result, it leads to PCIe bottleneck in multi-GPU environment. 

\textbf{KV cache compression} aims to reduce the size of the KV cache to improve the system throughput and support longer prompts. For example, tailored from the tensor compression techniques in DNN~\cite{oseledets2011tensor, chen2021re}, one can quantize into smaller data types~\cite{hooper2024kvquant,liu2024kivi,zhao2024atom,sheng2023flexgen} or sparsify the KV cache~\cite{kitaev2019reformer,10609626,tang2024quest}. Additionally, H2O~\cite{zhang2023h2o} and Scissorhands~\cite{liu2024scissorhands} retain only pivotal tokens in the KV cache, achieving significant size reductions without substantial performance degradation. In addition, ICAE~\cite{gecontext} and CacheGen~\cite{liu2023cachegen} introduce auto-encoding to transform the KV cache into an intact representation for storage and recover it during computation. While our {\name} is orthogonal to these works, it can complement them by compressing the KV cache before transfer, thereby reducing network traffic and improving transfer efficiency.

In summary, our {\name} focuses on accelerating the communication between prefill and decode workers as well as improving GPU resource utilization. Correspondingly, our techniques also differ from traditional disaggregated inference efforts: (i) While traditional efforts rely on message-passing-based communication, we introduce a tensor-centric communication design. (ii) With a dramatically shortened communication time consumption, we introduce a pull-based paradigm for KV cache communication while existing efforts lean toward a push-based pipelined approach.

\section{Discussion and Future Work}\label{sec:discussion}

We observe the underutilization of GPU memory on the prefill worker, especially with very long prompts, such as those exceeding 128K tokens. After the long prompt is granted GPU memory, it starts the prefill computation which will consume substantial space for the intermediate results, in addition to the KV cache. While the remaining memory space is considerable, it is insufficient for another long prompt. As a result, the remaining GPU memory is left idling. To address this, we plan to use the idling memory as a prefix cache.

Computational resources are sometimes underutilized in {\name}, particularly when there is an imbalance between prefill and decode workloads. In this case, the worker must wait for other workers to be available. To address this, rather than adding more nodes—a solution limited by potential GPU shortages and model loading delays—we propose to occasionally switch the role of prefill and decode workers. 
For example, if a request lingers in the KV cache queue on a decode worker, it could be redirected to a prefill worker with spare capacity. This prefill worker can serve as a helper for this decode worker. 
Likewise, a decode worker could temporarily run prefill tasks to better utilize available resources.

\section{Conclusion}
\label{sec:conclusion}
We introduced {\name}, a novel system designed to optimize disaggregated LLM inference by focusing on KV cache transfer. {\name} propose a novel tensor communication design, the associated system, and pull-mode transfer patterns. We implement {\name} and evaluate its performance across various workloads. Our findings confirm that {\name} reduces the per-request latency by 55\% compared to the baseline system under the same per-node QPS.




\bibliographystyle{plain}
\bibliography{references}

\end{document}